\documentstyle[aps,preprint]{revtex}
\newcommand{\journal}[4]{{{\sl #1}} {\bf #2}, {#3} (#4)}
\newcommand{\mprl}[3]{\journal{Phys.~Rev.~Lett.~}{#1}{#2}{#3}}
\newcommand{\mprb}[3]{\journal{Phys.~Rev.~B}{#1}{#2}{#3}}
\newcommand{\mpra}[3]{\journal{Phys.~Rev.~A}{#1}{#2}{#3}}

\newcommand{\cutau}{Cu$_{\rm 3}$Au}

\def\C44{C_{\rm 44}}
\begin{document}
\title
{Microscopic Modeling of the Growth of Order in an Alloy:
Nucleated and Continuous Ordering}
\author{Bulbul Chakraborty and Zhigang Xi\cite{Xipresent}}
\address
{Martin Fisher School of Physics\\
Brandeis University\\
Waltham, MA 02254, USA}

\maketitle
\begin{abstract}

We study the early-stages of ordering in \cutau\
using a model Hamiltonian derived from the effective medium theory of
cohesion in metals:  an approach providing a microscopic description of
interatomic interactions in alloys.
Our simulations show a crossover from a
nucleated growth regime to a region where the ordering does not follow
any simple growth laws.  This mirrors the experimental observations in
\cutau. The kinetics of growth, obtained from the simulations, is in
semi-quantitative agreement with experiments.
The real-space structures observed in our simulations offer some insight
into the nature of early-stage kinetics.
\end{abstract}
\narrowtext

The field of growth kinetics is concerned with understanding the
evolution of an initially disordered, high-symmetry phase into a final
equilibrium ordered state in which the symmetry is broken\cite{Bray}.
In alloys, the ordering process determines the microstructure  which
controls their
mechanical and electrical properties.  Simple, binary alloys such as \cutau,
also provide convenient testing grounds for theoretical models of ordering
kinetics\cite{Ludwig}.
When quenched from a high temperature disordered state to a temperature
below the order-disorder transition temperature,  alloys evolve from an
initially
metastable or unstable state\cite{Gunbook} towards the stable ordered
state.
The ordering process can be nucleated  or continuous and the
characteristics of early-stage growth are different for the two.   In
meanfield theory,
there is a well defined spinodal temperature at which the disordered phase
becomes unstable and continuous ordering sets in\cite{Gunton}.   The
linear, Cahn-Hilliard-Cook (CHC) theory of continuous ordering predicts
exponential growth of the structure factor  (at the ordering vector) below
the spinodal temperature and exponential relaxation above
it\cite{Gunton,Binder,Klein}.

The theoretical study of growth kinetics in alloys have
been based upon simulations of kinetic Ising models or simple Langevin
dynamics defined
by the Time-dependent Ginzburg-Landau (TDGL) model\cite{Bray,Gunton}.
These are the simplest models describing growth of order and, as such,
are essential for a fundamental understanding of these nonlinear
phenomena.  Simulations of the kind described in this work, based on
realistic interatomic interactions, bridge the
gap between Ising model simulations and experiments on real, metallic
alloys.  They also offer the opportunity of making quantitative comparisons
between theory and experiment and  the ability to distinguish the
universal characteristics of growth from the
system specific features.  In this paper, we compare results of simulations of
growth in \cutau\ to experimental results.  This is the first time that such a
quantitative comparison has been made between theory and experiments.

Experiments in \cutau\ have led to some interesting observations regarding the
crossover from  nucleated to continuous ordering.  The experimental
observations indicate two different temperature regimes; (a) a regime of
exponential relaxation, presumably followed by nucleated ordering and
(b) a regime which cannot be described either by an exponential
relaxation mechanism or by exponential growth\cite{Ludwig}.  The
crossover occurs at a temperature higher than the spinodal
temperature
(deduced from the divergence of the scattering intensity).  There seems
to be, therefore, a region in which the CHC model is not applicable.
Recent simulations
of a long-range Ising model\cite{Ngross} have led to the conclusion that the
linear theory of CHC  has limited applicability to early-stage growth kinetics.
In this work, we address these issues by simulating the kinetics of
ordering in
\cutau\ using a realistic model Hamiltonian.  The model is
derived from the Effective
Medium Theory of cohesion in metals (EMT)\cite{dth} which belong to the
general category of embedded atom approaches\cite{embedded}.

The results of our simulations bear a remarkable similarity to the
experimental
observations: (a) we observe a crossover behavior of exactly the same kind
as seen in experiments and (b) we deduce a spinodal temperature which is in
extremely good agreement with experiment.  This, in itself,
is an important set
of results since it
demonstrates the feasibility of making predictions regarding the microstructure
and early-stage growth kinetics of alloys starting from  a model based
on electronic structure information.  With simulations, however, one can
go a step further and attempt to
understand the reasons behind these growth
processes by probing the system at various length scales.  We have taken
a step in that direction, in this paper, by correlating real-space
structures with the growth of the structure factor and by comparing them to
that expected from pure linear theory
and to structures seen in simulations of long-range Ising models\cite{Ngross}.

The model Hamiltonian used in this study
has been described in
detail elsewhere\cite{jphys,bcprl,xithesis}.  It has been shown that the model
provides an excellent description of the equilibrium statistical mechanics of
Cu-Au alloys\cite{jphys,bcprl}.
The current study extends the application of this model to
phenomena occurring far from equilibrium.
The late-stage
growth has been investigated in an earlier work\cite{Xi} which confirmed the
existence of anisotropic scaling and predicted an anisotropy factor in
excellent agreement with experiment\cite{Nagler}.   One of the features
of atomistic
models is that they can predict system-specific features of this kind.
The late-stage of ordering kinetics which describes the coarsening process, is
controlled by the motion of interfaces separating two different ordered
domains\cite{Bray,Gunton}.  In contrast, the early-stage
growth depends on the free energy differences between the ordered and
disordered phases\cite{Gunton,Binder} and by extending our study to
this regime we are not only
exploring new phenomena but also analyzing  new areas of applicability of the
EMT model.

We perform Monte Carlo simulations which allow for exchanges of atoms and
changes in shape and volume of the shape of the simulation box.   This
accommodates
any homogeneous strain accompanying the order-disorder transition.   For a
completely, realistic simulation of the ordering process, local displacements
of atoms should also be included.   However, for the \cutau\ ordering, these
are not expected  play a major role since the ordered phase has the same
crystalline symmetry as the high-temperature disordered phase.

Of the simple ordering  alloys, \cutau\ is one of the more interesting ones
even
when described by a kinetic Ising model.  The groundstate of this model is
four-fold degenerate and TDGL
models based on phenomenology and symmetry arguments lead to
a three component order parameter with anisotropic gradient terms\cite{Lai}.
We
are unaware of any simulations of the kinetic Ising model appropriate to
\cutau,
but the results of our simulations of late stage growth were found to be
in essential agreement with the study of late stage growth in
the TDGL model\cite{Lai}.

The predictions of the TDGL model regarding early-stage growth kinetics is
different from the CHC predictions only through the appearance of an
anisotropic relaxation time arising from the unusual gradient terms.  The
linear approximation for \cutau\ leads to exponential growth or relaxation
with a growth rate given by:
\begin{equation}
\label{Dq}
D_{\bf q} = -M(r (T - T_{sp}) + C_1 (q_{||})^2 + C_2 (q_{\perp})^2 )  ~.
\end{equation}
Here, $M$ is the mobility and $r$ the coefficient of the quadratic term
in Landau theory.
$T_{sp}$ defines the classical spinodal temperature, and the
coefficients $C_1$ and $C_2$ reflect the anisotropy of the gradient
terms.  The anisotropy factor, the ratio of $C_1$ to $C_2$, in \cutau\
is predicted to be $\simeq 2.5$ from EMT and this value is in good
agreement with experiments\cite{Nagler}.   The wavevector $q$ in Eq.
(\ref{Dq}) is measured from the superlattice Bragg peak and $q_{||}$ and
$q_{\perp}$ denote the radial and transverse components.
The relaxation time, $\tau ({\bf q})$,
is given by the inverse of $D({\bf q})$, and the structure factor grows
according
to\cite{Klein}:
\begin{equation}
\label{struc}
S ({\bf q}) = S_0 \exp (2D({\bf q}) t) + ({k_B T}/2D({\bf q})) (\exp (2D({\bf
q})t) - 1) ~.
\end{equation}
The second term,
referred to commonly as the Cook term, arises from averaging over thermal
noise\cite{Klein}.
Because of the anisotropy, the structure factor grows at different rates
along the radial and transverse directions.  In the current work, we
concentrate on studying the averages of the structure factor over the radial
and
transverse directions which are more appropriate for comparing to
experiments on powder samples.

As can be seen from Eq. (\ref{Dq}), linear theory would predict  a
positive value of $\tau$ (exponential growth) for the superlattice peak
at temperatures below $T_{sp}$ and a negative value (exponential
relaxation) for temperatures above $T_{sp}$.  The limiting value of the
intensity
in the relaxation regime is predicted by Eq. (2) to be ${k_B T}/(2D({\bf q}))$
which is seen to diverge at $T_{sp}$,  as does the relaxation time.
The simulation results are
to be compared to these predictions.

The simulations are carried out in a box with a
linear dimension of 30 times the
lattice parameter along each direction and containing 108,000 atoms.  The alloy
is annealed at a temperature of $\approx 750K$ and quenched to
temperatures ranging from approximately 20K above the transition temperature,
$T_{tr}$
(642K from our simulations), to approximately 30K  below. At each
quench temperature, the structure factor at the superlattice Bragg peaks
and at the closest points to these peaks (at a distance of $2\pi /30$
because of our periodic boundary conditions) are calculated by averaging
over 10 independent runs.  We performed two sets of simulations, one
with a fixed lattice and another where the three lattice parameters were
treated as Monte Carlo variables.  It was found that the lattice relaxed
very quickly and did not influence the growth kinetics of the chemical
order.  This is understandable for the \cutau\ alloy where the chemical
short-range order drives the change in volume and there are no elastic
strains arising from misfits between the disordered and ordered
regions
which have the same crystalline symmetry.

The growth of the structure factor at various quench temperatures is
shown in Fig. (\ref{fig:1}).  The structure factors shown are averages
over the three equivalent superlattice peaks.  In addition, we have
folded in  the values at the set of $q$ vectors closest to the Bragg
peak to improve our statistics and simulate a finite q-space resolution.
The growth of the isolated superlattice peak show the same trends but is
noisier because of poorer statistics.

The structure factor shows clear exponential relaxation in the
temperature range between 636K and 660K.  As expected, there is no
change in going across the first-order transition in this nucleated
growth regime.  As discussed earlier, the CHC linear theory
predicts that the relaxation time, $\tau$, diverges at the classical
spinodal as does the stable (or metastable) intensity at the
superlattice point.  We have fitted our simulation results to the CHC
predictions for the exponential relaxation regime and the results are
shown in Fig. (\ref{fig:2}).  By extrapolating the inverse of the
limiting intensity at the superlattice peak, $I_{infty}$, we determine
a
classical spinodal temperature of 615K, which is 27K below the
transition temperature; in excellent agreement with experiment\cite{Ludwig}.
We
find that as we approach the classical spinodal temperature,
the simulation data can no longer be described by simple exponential
relaxation.
In
addition, we find that the results below the classical spinodal
temperature cannot be described by simple exponential growth.  There
seems to be a range of temperature, around the classical spinodal
temperature,
where the simulations
results do not agree with the CHC predictions.  These observations
mirror the experimental observations in \cutau\cite{Ludwig}.

We have also compared the fluctuation relaxation times, $\tau$, obtained from
our
simulations with the experimental values.  This comparison requires a mapping
of
our Monte Carlo timescale to the actual experimental timescales in \cutau.  We
performed this mapping by using the known experimental value of the activation
barrier in \cutau\cite{Barrier}, which generates the temperature dependence of
the timescale, and fitting the simulation relaxation time to the experimental
relaxation time at the highest quench temperature.
The experimental and the theoretical values are compared in Table 1.  Both
experiments and simulations indicate that the
relaxation time diverges as ${1} / {|T - T_{sp}|^{1.4}}$.
Using this mapping of time scales, the simulated and experimental results for
the
time-dependence of the structure factor can be compared in detail.  An example
of
this comparison is shown in the inset in Fig. 1.  The temperature of comparison
is just above the classical spinodal temperature, where the data start
deviating
from the purely relaxational regime, and much below the temperature used for
fitting the relaxation time.

The excellent agreement between theory and experiment shows that the EMT
interactions provide a very good description of the dynamics in these alloys.
It
should be emphasized, that the EMT model is a microscopic model which does not
rely upon fitting to any aspect of the phase diagram of these alloys.  The
parameters in the model are obtained from
fitting to ground-state properties such as the formation energy of the 50-50
alloy and the bulk moduli of pure $Cu$ and pure $Au$\cite{jphys,bcprl}.
We have performed a set
of simulations where the alloy was quenched from a very high temperature
and there is no qualitative difference in the results.

The simulations lead to two important conclusions at this point: (i) the
experimental observations are reproduced, semi-quantitatively,
by a simulation based on simple
atom-exchange dynamics using a realistic model Hamiltonian,
and (ii) The initial stages of ordering show
features which are different from those predicted by the simplest theory
of early-stage kinetics, the CHC model.

At temperatures slightly below the classical spinodal temperature, the
simulations indicate a crossover from a purely relaxational regime to a growth
regime as a function of time.  At the shortest of time scales, the system
behaves
as it would if the disordered state with unbroken symmetry,
was stable or metastable and relaxes towards
this minimum. At later times, the order grows as the system evolves towards the
ordered phases.  This later phase is absent at temperatures well above the
spinodal temperature where the initial state is clearly stable (metastable).
The
simulations, therefore, show an early-stage evolution preceding the growth
regime
predicted by the CHC linear theory.  An interesting question is what sets the
timescale for crossover from the relaxational to the growth regime.  It would
be
expected that this timescale gets shorter as the temperature is lowered below
the classical spinodal.

We have tried to understand the initial ordering process by examining
the real space structures at different temperatures.  In Fig 3., we show
snapshots from two different temperatures: (a) $T = 636K$, at which
there is clear exponential relaxation and we are in the nucleated growth
regime, and (b) $T = 612K$, which is slightly below the classical
spinodal temperature and clearly outside the exponential relaxation
regime.    To generate the snapshots, the local order parameters were
calculated
by coarse graining over one cubic unit cell.  If the value of the local order
parameter was $\simeq 1$, a symbol was generated.  The
four different symbols correspond to the four possible ordered domains or,
equivalently, the four types of sublattice ordering.   The snapshots,
therefore,
show areas where the order parameter is large.  They do not show
long-wavelength
fluctuations of small magnitude.

The feature to be noticed in the snapshots is that at both temperatures, tiny
ordered domains are visible at the earliest times and there is no
qualitative difference in the morphology at the two temperatures.
These snapshots are
consistent with the behavior of the structure factors which show
exponential relaxation at the earliest times for the full range of
temperatures studied here and observed in experiments.   At later times,
the structure factors show two distinctly different behaviors.  For
temperature $T \geq 636K$, there is pure relaxation and below this
temperature, there is growth of order.  Observing the real-space
structures at times close to where the deviation becomes evident, we find
that at $T= 636K$ (pure relaxation), the ordered domains are still
well separated.  However, at $T = 612K$, the ordered domains start
to interact, around this time, and show an interconnected structure
which extends throughout the sample.  The two dimensional cuts shown in
Fig. 3 are suggestive but a complete three-dimensional mapping is needed
to see the structure.  We have performed an approximate
three-dimensional map which shows this picture.
The cuts show clearly that
the morphology at the two different temperatures start looking distinct
at around the same time where the structure factors deviate from one
another.   This is a qualitative picture, and to make quantitative predictions,
techniques for identifying connected clusters will have to be used.

The picture that seems to be emerging is very similar to the
one that followed the analysis of continuous ordering in two-dimensional,
long-range Ising
models\cite{Ngross}.  The ordering begins at small length scales with tiny
domains of the different ordered phases.  As these domains grow, they
start interacting with each other and at some timescale, determined by
the quench
temperature, they form a percolating ordered path through the sample.
Correlating the behavior of the structure factor with the
real-space structures, we seem to find that at this time the nature of
the ordering
fluctuations changes from relaxation to growth.
It is
remarkable that the microscopic model exhibits a growth morphology
akin to that
observed in the Ising models.  The similarity provides an avenue for
analyzing our real-space structures by extending the tools developed for
Ising models, {\it ie} through the mapping to percolation clusters\cite{Perc}.
This
should lead to a more rigorous description of the relationship between
real-space structures and early-stage growth kinetics.

After the completion of this work, we came across an analysis of the TDGL
model which shows the same crossover from
relaxation to growth\cite{zanneti}.  It seems therefore, that the kinetics of
ordering in \cutau\ can be understood within the framework of the TDGL models
and
the particular parameters characterizing this alloy are such that the initial
relaxational regime is observable in experiments and simulations.

To conclude, the simulations based on the microscopic,
EMT hamiltonian have led to an interesting picture of early-stage growth
in alloys.  The microscopic model bridges the gap between
experiments on real alloys and kinetic Ising models.  The results of the
microscopic model are in excellent agreement with experiment.  These
simulations have, for the first time, provided the opportunity of making
a quantitative comparison between theoretical models and experiment and
by making a connection to Ising model simulations, provided us with an
opportunity to construct a theory of kinetics of growth in real,
metallic alloys.

We would like to thank W. Klein and K. F. Ludwig for many enlightning
discussions.  We gratefully acknowledge the support of NSF through the grant
DMR-9208084.

\begin{table}
\caption{Fluctuation relaxation times obtained from our simulations are
compared
to the values extracted from experiments (Ref. 2).  The experimental values
were
read off from the graph in Ref. 2 and the simulation result was fitted to the
experimental value of $\tau$ at $T = 1.028 T_{tr}$.}
\begin{tabular}{llll}
&$T/T_{tr}$&Experimental $\tau$ (s)&Theoretical $\tau$ (s)\\
    \tableline
&1.028&0.5&0.5\\
&1.009&1.0&1.1\\
&0.991&4.0&4.0\\
&0.972&13.0&13.4\\
\tableline
\end{tabular}
\label{tab:one}
\end{table}

\begin{figure}
\caption{ The structure factor as a function of time is shown over a range of
quench temperatures.  Starting from the lowest curve,
the quench temperatures are
$T/T_{tr} = 1.028, 1.009, 0.991, 0.972$, and $0.953$.
The transition temperature ($T_{tr}$) is 642K.  The inset shows a comparison of
the
experimental data and simulation data at $T/T_{tr} = 0.972$.
The squares are the
experimental points obtained from Ref. 2 and the dots are results of our
simulations.}
\label{fig:1}
\end{figure}

\begin{figure}
\caption{The inverse of the
limiting intensity at the superlattice Bragg peak, $I_{infty}$, obtained by
fitting the structure factors
to an exponential relaxation form, is plotted as a function of temperature.
The
classical spinodal temperature, defined to be the point where this intensity
diverges, is found to be 615K.}
\label{fig:2}
\end{figure}

\begin{figure}
\caption{Real-space structures at 612K (left panel) and 636K (right panel).
The plots show two dimensional
cuts from a particular Monte Carlo run.  The structures start looking
qualitatively different for the two different temperatures at approximately 20
Monte Carlo steps.}
\label{fig:3}
\end{figure}

\begin{references}
\bibitem[*]{Xipresent}Present Address: Lotus Corporation, Westford, MA.
\bibitem{Bray} A. J. Bray,  \journal{Physica}{A194}{41}{193}, and to
appear in {\it Advances in Physics}, 1995.
\bibitem{Ludwig} K. F. Ludwig {\it et al}, \mprl{61}{1859}{1988}
\bibitem{Gunbook} J. D. Gunton and M. Droz, {\em Introduction to the theory of
metastable and unstable states}, Lecture Notes in Physics, Vol. 183,
(Springer-Verlag, Berlin-Heidelberg, 1983)
\bibitem{Gunton}
J. D. Gunton, M. San Miguel and Paramdeep S. Sahni, in {\em
Phase Transitions and Critical Phenomena,} Vol. 8, eds. C. Domb and J. L.
Lebowitz, (Academic Press, London, 1983)
\bibitem{Binder}K. Binder, \mpra{29}{341}{1984}.
\bibitem{Klein}W. Klein and G. Batrouni, \mprl{67}{1278}{1991}.
\bibitem{Ngross} N. Gross, W. Klein and K. F. Ludwig, \mprl{73}{2639}{94}.
\bibitem{dth} J. K. N\o rskov, K. W. Jacobsen and M. J. Puska,
\mprb{35}{7423}{1987}
\bibitem{embedded} M. S. Daw, S. M. Foiles and M. I. Baskes, \journal{Mat.
Sci. Reports}{9}{251}{1993}
\bibitem{jphys} Zhigang Xi {\it et al}, \journal{J. Phys: Condens
Matter}{4}{7191}{1992}
\bibitem{bcprl} Bulbul Chakraborty and Zhigang Xi, \mprl{68}{2039}{1992}
\bibitem{xithesis} Zhigang Xi, Brandeis University thesis, unpublished
\bibitem{Xi} Zhigang Xi and Bulbul Chakraborty, \journal{Mat. Res. Soc.
Symp. Proc.}{291}{165}{1993}
\bibitem{Nagler} S. E. Nagler {\it et al}, \mprl{61}{718}{1988}
\bibitem{Lai} Z. W. Lai, \mprb{41}{9239}{1990}
\bibitem{Barrier} Daniel B. Butrymowicz, John. R. Manning and Michael E. Read,
\journal{J. Phys. Chem. Ref. Data}{3}{527}{1974}.
\bibitem{Perc}W. Klein, \mprl{65}{1462}{1990}.
\bibitem{zanneti}F. Corberi, A. Coniglio and M. Zannetti, preprint.
\end{references}
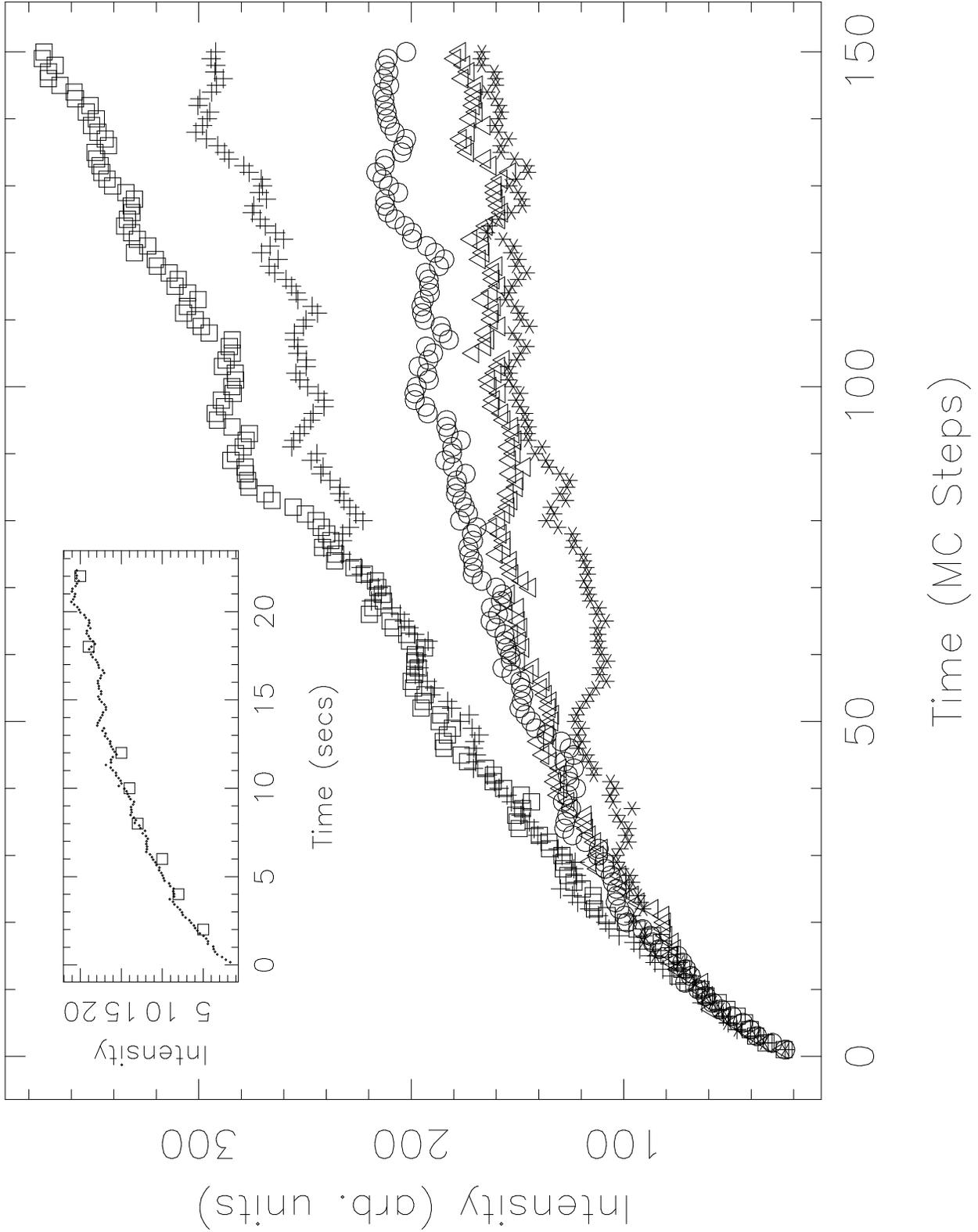
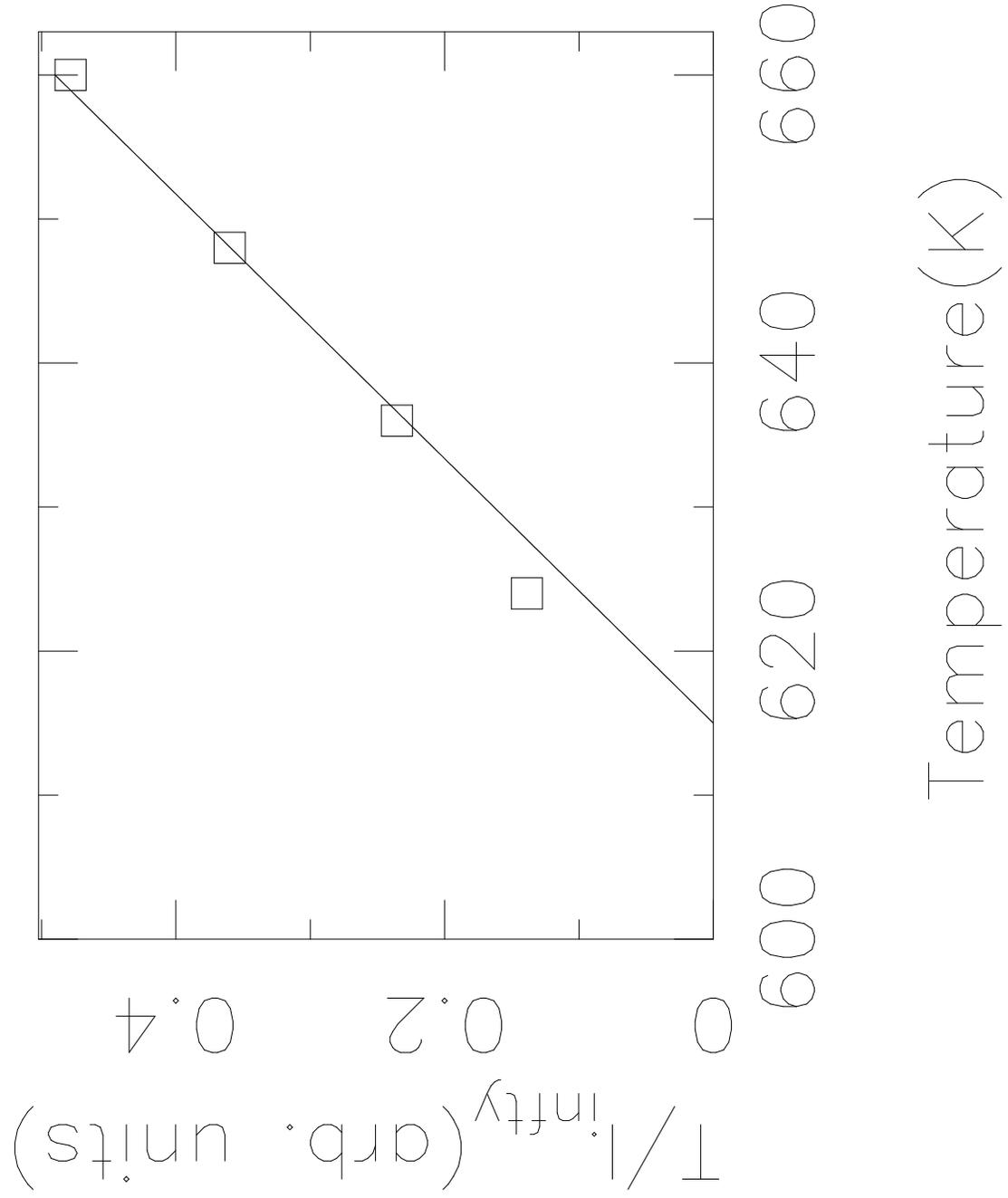
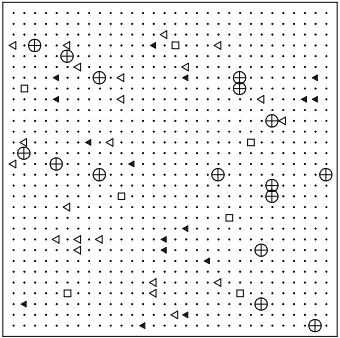
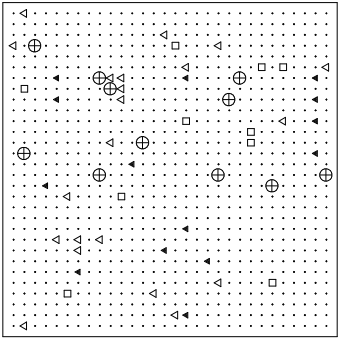
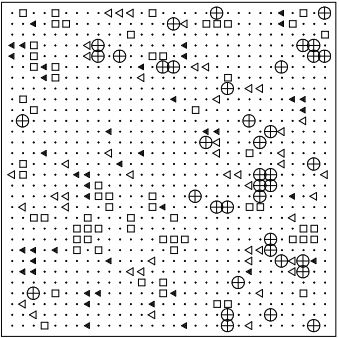
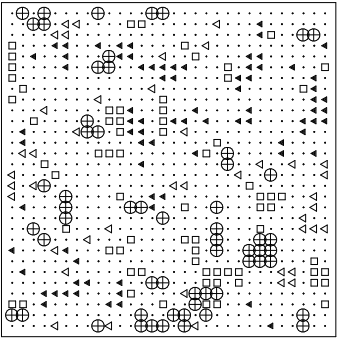
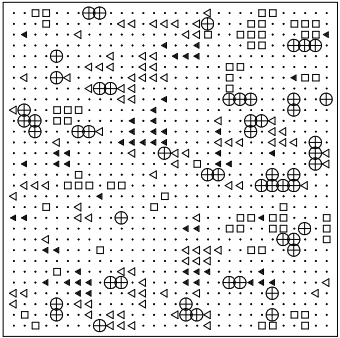
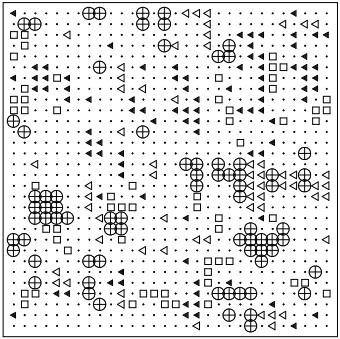
\end{document}